\documentclass[aps,prd,twocolumn, superscriptaddress,floatfix, nofootinbib, noshowpacs,preprintnumbers]{revtex4-2}
\usepackage{lipsum} 
\usepackage[utf8]{inputenc}
\usepackage[english]{babel}
\usepackage[colorlinks=true,citecolor=blue]{hyperref}
\usepackage{amsmath}
\usepackage{amsfonts}
\usepackage{numprint}
\usepackage{amssymb}
\usepackage{graphicx}
\usepackage{siunitx}
\usepackage{natbib}
\usepackage{url}
\usepackage{placeins}

\DeclareSIUnit\barn{b}

\begin{document}

\title{Learning to see R-parity violating scalar top decays}

\author{Gerrit Bickendorf} 
\email[]{bickendorf@th.physik.uni-bonn.de} 
\affiliation{Bethe Center for Theoretical Physics and Physikalisches Institut,
  University of Bonn, Bonn, Germany }

\author{Manuel Drees} 
\email[]{drees@th.physik.uni-bonn.de} 
\affiliation{Bethe Center for Theoretical Physics and Physikalisches Institut,
  University of Bonn, Bonn, Germany } 

\begin{abstract}
  With this article we introduce recent, improved machine learning methods from
  computer vision to the problem of event classification in particle
  physics. Supersymmetric scalar top decays to top quarks and weak
  scale bino-like neutralinos, where the neutralinos decay via the
  $UDD$ operator to three quarks, are difficult to search for and
  therefore weakly constrained. The jet substructure of the boosted
  decay products can be used to differentiate signal from background
  events. We apply transformer-based computer vision models CoAtNet
  and MaxViT to images built from jet constituents and compare the
  classification performance to a more classical convolutional neural
  network (CNN). We find that results from computer vision translate
  well onto physics applications and both transformer-based models
  perform better than the CNN. By replacing the CNN with MaxViT we
  find an improvement of $S/\sqrt{B}$ by a factor of almost 2 for some
  neutralino masses. We show that combining this classifier with
  additional features results in a strong separation of background and
  signal. We also find that replacing a CNN with a MaxViT model in a
  simple mock analysis can push the 95\% C.L. exclusion limit of stop
  masses by about $100$ GeV and $60$ GeV for neutralino masses of
  $100$ GeV and $500$ GeV.
\end{abstract} 
 
\maketitle 

%\tableofcontents
\section{Introduction}

The minimal supersymmetric extension of the standard model (MSSM) is a
promising candidate for physics beyond the standard model
\cite{WESS197439, NILLES19841, HABER198575, FAYET1977249,
  doi:10.1142/4001, MARTIN_1998} that might solve the hierarchy
problem. Despite many experimental searches, most notably by the ATLAS
and CMS collaborations at the CERN LHC in recent years, no conclusive
evidence of its realization in nature has been found, pushing the
parameter space to ever higher masses (see ref. \cite{Workman:2022ynf}
for an overview). In models with conserved $R-$parity (RPC), many
searches leverage the large $p_T^\text{miss}$ due to the stable
lightest supersymmetric particle (LSP) leaving the experiment
undetected \cite{atlascollaboration2024quest}.  Once the RPC
assumption is dropped, these strategies often become insensitive. In
the context of the $R-$parity violating (RPV) MSSM, new terms are added
to the superpotential that break lepton- or baryon-number conservation
\cite{Barbier_2005}. These additional terms imply that drastically
different search strategies are needed \cite{redelbach2015searches},
especially when prompt decays of supersymmetric particles become
drowned by the hadronic activity inside the detector.

At a hadron collider like the LHC, the production of strongly
interacting superparticles has the largest cross section for a given
mass. Among these, the stops -- the superpartners of the top quark --
are often assumed to be the lightest. On the one hand, for equal
squark masses at some very high (e.g. Grand Unified or Planckian)
energy scale, renormalization group effects reduce the masses of the
stops; mixing between the masses of the $SU(2)$ doublet and singlet
stops will reduce the mass of the lightest eigenstate even more
\cite{MARTIN_1998}. On the other hand, simple naturalness arguments
\cite{Feng:1999mn, Kitano:2006gv, Papucci:2011wy} prefer not too heavy
stop squarks, but allow much heavier first and second generation
squarks. This motivates the analysis of scenarios where the mass of
the lighter stop squark lies well below those of the other strongly
interacting superparticles.

The same naturalness arguments also prefer rather small supersymmetric
contributions to the masses of the Higgs bosons. In most (though not
all \cite{Ross:2017kjc}) versions of the MSSM this implies rather
light higgsinos, typically below the stop. Since the mass splitting
between the three higgsino-like mass eigenstates is small, they all
behave similarly if the LSP is higgsino-like. In particular, in the
kind of RPV scenario we consider, all three states would lead to very
similar ``fat jets'' when produced in stop decays; the recognition of
such jets by exploiting recent developments in computer vision is one
of the central points of our paper, which would apply equally to all
three higgsino states. However, about half of all stop decays would
then produce a bottom, rather than a top, together with a higgsino,
thereby complicating the analysis of the remainder of the final state.
Moreover, higgsinos being $SU(2)$ doublets have a sizable direct
production rate. Their non-observation therefore leads to significant
constraints on parameter space, especially (but not only) if the bino
has mass comparable to or smaller than the higgsinos \cite{Baer:2020sgm,
ATLAS:2023lfr, ATLAS:2024umc, ATLAS:2024tqe, CMS:2022vpy, CMS:2024gyw}.

In order to avoid such complications, we consider the pair production
of scalar top quarks which decay to top quarks plus two neutralinos
with unit branching ratio. The neutralinos in turn decay promptly via
the $UDD$ $R-$parity breaking term which is fairly difficult to
constrain \cite{Evans_2013}. Since each neutralino may form a (fat)
jet one can use the substructure to differentiate it from background
processes \cite{Butterworth_2009}. One may represent the jets as
images made of calorimeter cell hits that can be used by computer
vision techniques. Using a convolutional neural net (CNN) has already
been shown to work well on these images \citep{Macaluso_2018,
  ATL-PHYS-PUB-2017-017, plehn2022modern, Komiske_2017,
  Kasieczka_2017, ATL-PHYS-PUB-2023-001, Sirunyan_2020,
  PhysRevD.106.055008, PhysRevD.98.076017, Lee_2019,
  filipek2021identifying, han2023guide}.

In recent years computer vision techniques have improved drastically
with novel approaches such as the vision transformer
\cite{dosovitskiy2021image}. In standardized computer vision tasks,
these models have been shown to outperform CNN-based models for large
data sets. Fortunately generating large sets of simulated events is
relatively cheap in particle physics which motivates the use of these
new techniques. Transformers have already been applied to classification in particle
physics scenarios \cite{Mikuni_2021,Di_Bello_2023,builtjes2024attention,hammad2024multiscale,qu2024particle,Finke_2023,he2023quarkgluon,hammad2024streamlined,hammad2024exploring}, although these focus on representing the jet as a set of particles, instead of as an image.

In this article, we for the first time apply two
modern transformer-based computer vision techniques to find
neutralinos from scalar top quark decays and compare the results to a
classical CNN to see if the gain in performance translates to detector
images. Using gradient-boosted decision trees (GBDT), we combine the data from
both neutralinos tagged in this way and add further high-level
features to construct our final event classifier.

The remainder of this article is structured as follows: In section
\ref{sec:signal} we describe the specifics of the signal model we
use. In section \ref{sec:datagen_preselection} we show how we
generated the data sets to which we apply the preprocessing outlined
in section \ref{sec:preprocessing}. The novel computer vision
architectures we wish to adapt are described in section
\ref{sec:architectures}. Section \ref{sec:dataset_creation} outlines
the generation of datasets from neutralino decay and background events, which are used to train the
neutralino taggers in section \ref{sec:train_base}. The performance of
these taggers is discussed in section \ref{sec:res_base_classifiers}.
In section \ref{sec:boosted_classifiers} we combine information from
both tagged neutralinos into even more powerful classifiers. We also
demonstrate the power of combining the neutralino taggers with other
high-level features in section \ref{sec:add_high_level_features}. The
improved stop mass reach is shown in section \ref{sec:application_137},
while section \ref{sec:conclusion} contains a brief summary and some
conclusions.

\section{Signal model}
\label{sec:signal}

\begin{figure}[h]
    \centering
    \includegraphics[width=0.4\textwidth]{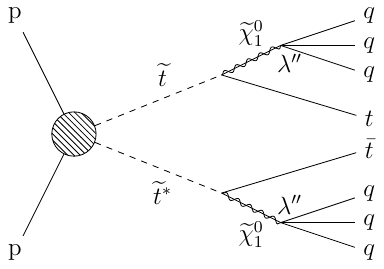}
    \caption{Stop pair production with each stop decaying to a top quark
      and a neutralino. The neutralinos decay via the RPV $UDD$ operator
      with nonzero $\lambda''$.}
    \label{fig:feynman}
\end{figure}

The MSSM contains two scalar top quarks which are mixtures of the
$SU(2)$ doublet $\tilde{t}_L$ and singlet $\tilde{t}_R$ weak gauge
eigenstates. We will work with breaking parameters such that the
lighter mass eigenstate $\tilde{t}_1$ contains mainly the right-handed
top squark which decays promptly into a top quark and the bino-like
neutralino $\tilde{\chi}_1^0 \equiv \tilde\chi$. We consider all other
scalar quarks to be decoupled. In order to avoid the constraints from
missing $E_T$ based searches we add
\begin{equation} \label{WRPV}
W_{\not{R}_p} = \frac{1}{2}\lambda''_{ijk}U_i^cD_j^cD_k^c
\end{equation}
to the superpotential, where $U_i^C$ and $D_i^c$ are the up and down
type $SU(2)$ singlet chiral quark superfields and $i,j,k$ are
generation indices. Clearly, this term violates baryon number
conservation. In eq.(\ref{WRPV}) anti-symmetrization over color (i.e.
contraction with the totally antisymmetric tensor in color space) is
implied, hence the coupling has to be antisymmetric in the last two
indices. Therefore there are in general $9$ independent coupling
constants $\lambda''_{ijk}$. When $i=3$ this would allow the stop to
decay directly to two lighter quarks which has already been
extensively studied \cite{201598, Aad_2016, ATLAS:2017jnp,
  PhysRevD.98.112014}. A coupling with $i \neq 3$ allows even a light
neutralino to decay into three quark jets via an off-shell squark.
The process we are interested in is shown in figure \ref{fig:feynman}.

We also note that a mostly $\tilde t_R$ eigenstate decaying into a
bino-like neutralino produces a predominantly right-handed top
quark. The same is true for a $\tilde t_L$ decaying into a neutral
higgsino. In contrast, a $\tilde t_L$ decaying into a bino or a
$\tilde t_R$ decaying into a neutral higgsino would produce a
right-handed top quark. Since we do not try to reconstruct the
polarization of the top (anti)quark in the final state, all four
reactions would have very similar signatures, and could be treated
with the methods developed in this paper. However, as already noted in
the Introduction, a light neutral higgsino implies the existence of a
nearly mass degenerate charged higgsino (and of a second neutral
higgsino), thereby reducing the branching ratio for
$\tilde t \rightarrow t + \tilde\chi$ decays. Moreover, by $SU(2)$
invariance a mostly $\tilde t_L$ stop eigenstate would be close in
mass to $\tilde b_L$, leading to additional signals from $\tilde b_L$
pair production. By focusing on a mostly $\tilde t_R$ lighter stop
and a bino-like LSP we avoid these complications.

\section{Data generation and preselection}
\label{sec:datagen_preselection}

For baseline selections, we follow roughly the CMS search for this
signal process \cite{PhysRevD.104.032006}. We impose the following
preselection cuts:
\begin{enumerate}
\item One muon with $p_T>\SI{30}{\giga \eV}$ or electron with
  $p_T>\SI{37}{\giga \eV}$ and $\vert \eta \vert < 2.4$. 
\item The lepton must be isolated within a cone radius depending on
  the $p_T$ of the lepton like
\begin{equation*}
   R=\begin{cases}
 0.2 & p_T <\SI{50}{\giga \eV}\\
 \SI{10}{\giga \eV} /p_T & \SI{50}{\giga \eV}<p_T < \SI{200}{\giga \eV}\\
        0.05 & p_T > \SI{200}{\giga \eV}
     \end{cases}
\end{equation*}
Together with the first cut, this isolation requirement implies that
in almost all events the lepton originates from the semileptonic
decay of one of the top (anti)quarks in the final state. These two
cuts satisfy the requirements of the single lepton trigger. Note that
the events must contain exactly one such isolated lepton; this largely
removes $Z +$ jets backgrounds.
\item We define ``AK04 jets'' via the anti-$k_T$ jet clustering
  algorithm with distance parameter $R=0.4$, requiring
  $p_T>\SI{30}{\giga \eV}$ and $\vert \eta \vert <2.4$ for each jet.
  We demand that the event contains at least $7$ such AK04 jets, at
  least one of which is $b$-tagged. We note that our signal events
  contain at least two $b$ (anti)quarks from top decay. Moreover, even
  if both $t$ and $\bar t$ decay semi-leptonically, signal events
  contain $8$ energetic quarks even in the absence of QCD radiation. They
  should therefore pass this cut with high efficiency, except for very
  light neutralinos where several of their decay products might end up
  in the same (quite narrow) AK04 jet. On the other hand, SM $t \bar t$
  events with one top decaying semi-leptonically contain only $4$ hard
  quarks. Hence at least three additional jets would have to be produced
  by QCD radiation, significantly reducing the $t \bar t$ background, and
  reducing the $W+$ jets background even more.
\item $H_T> \SI{300}{\giga \eV}$, where $H_T$ is the scalar sum of
  the transverse momenta of all AK04 jets. This cut is mostly effective
  against $W,Z+$ jets backgrounds.
\item At least one combination of $b$-tagged jet and isolated lepton
  must have an invariant mass between $\SI{50}{\giga \eV}$ and
  $\SI{250}{\giga \eV}$. Most events where the lepton and the $b$ quark
  originate from the decay of the same $t$ quark pass this cut, which
  helps to further reduce the $W+$ jets background. 
\item At least one AK08 jet (defined with distance parameter
  $R = 0.8$), with $p_T > \SI{100}{\giga \eV}$. We will later try to
  tag these ``fat jets'' as coming from neutralino decay. However, a
  boosted, hadronically decaying top (anti)quark can also produce such
  a jet. We will also consider even fatter jets. Since (nearly) all
  particles inside an AK08 jet will end up inside the same jet if
  $R>0.8$ is used in the jet clustering, while these fatter jets will
  contain additional ``nearby'' particles, they will automatically
  also have $p_T > \SI{100}{\giga \eV}$.
  
\end{enumerate}

After these cuts, the remaining background is almost exclusively due
to top quark pair production as can be seen in the original CMS
publication \cite{PhysRevD.104.032006}. In our simulation we therefore
only consider this background process.

For the signal model, we set the masses of squarks (except that of the
stop), gluinos, wino- and higgsino-like neutralinos to
$\SI{5}{\tera \eV}$. We only set one RPV coupling nonzero,
$\lambda''_{223}=-\lambda''_{232}=0.75$; this leads to prompt
neutralino $\tilde\chi \rightarrow csb$ decay even if the
exchanged squark has a mass of $5$ TeV,
$\tau_{\tilde \chi} \sim 10^{-18} \ {\rm s} \cdot [m_{\tilde
  \chi}/(100 \ {\rm GeV})]^{-5}$. We scan over the stop mass from
$m_{\widetilde{t}}=\SI{700}{\giga \eV}$ to
$m_{\widetilde{t}}=\SI{1200}{\giga \eV}$ in steps of
$\SI{25}{\giga \eV}$. We also scan over the neutralino mass from
$m_{\tilde\chi}=\SI{100}{\giga \eV}$ to
$m_{\tilde\chi}=\SI{500}{\giga \eV}$ in $\SI{10}{\giga \eV}$
steps.

Background and signal events are simulated using
\textsc{MadGraph5\_aMC@NLO 3.2.0} \cite{Alwall_2014}. The $t \bar t$
background is generated with between $0$ and $3$ additional matrix
element partons while the signal events contain up to 2 additional
partons. The NNPDF3.1 PDF-set \cite{Ball_2017} is used. We use
\textsc{pythia 8.306} \cite{bierlich2022comprehensive} for parton
showering and hadronization; background events are showered with the
CP5 tune while signal events are showered with the CP2 tune
\cite{pythia_tunes}. Events with different matrix element level final
state parton multiplicities are merged with the MLM prescription
\cite{Mangano_2007}, in order to avoid double counting events where
the parton shower produces additional jets. Finally, detector effects
are simulated with the CMS card of \textsc{Delphes 3.5.0}
\cite{de_Favereau_2014,Mertens_2015}.

\section{Preprocessing}
\label{sec:preprocessing}

The main novelty of this paper is the adoption of very recent computer
vision techniques to tag the hadronically decaying neutralinos. To
that end we first have to translate the simulated detector data to
images.

The objects we are interested in are jets clustered with the
anti-$k_T$ (AK) jet algorithm as implemented by the FASTJET package
\cite{Cacciari_2012}. Choosing the optimal distance parameter $R$ for a
given purpose can be somewhat nontrivial. A small value of $R$ means
that most particles inside a sufficiently hard jet originated from the
same parton, but some of the energy of that parton might not be
counted in this jet due to final state showering. On the other hand, a
large $R$ likely leads to jets that capture all daughter particles
while also muddying the waters by including unrelated objects,
e.g. from initial state showering. One can use the fact that the decay
products of a resonance with a fixed mass $m$ and transverse momentum
$p_T$ spread roughly like
$\Delta R = \sqrt{\Delta\phi^2+\Delta \eta^2}\propto m/p_T$ and the
typical energy scale of the process to arrive at a {\it best guess}
for an optimal $R$ parameter. This can be aided by the use of jet
clustering algorithms with variable $R$ (e.g.\cite{Chakraborty2022,
  Krohn_2009}). In the case at hand this optimal value of $R$ would
depend on both the stop and the neutralino mass. We therefore do not
work with a single fixed value of $R$, but instead we will cluster
each event using several values of $R$, and ensemble the resulting jet
images to get a better per-event classification. Because we consider
rather large neutralino masses,
$m_{\tilde\chi} \geq 100 {\rm GeV}$, we consider AK08
($R=0.8$), AK10 ($R=1.0$) and AK14 ($R=1.4$) jets. This also allows us
to keep the technique general, i.e. to use the same algorithm over the
entire parameter space. Recall that the resulting fat jet has to
satisfy $p_T > 100$ GeV and $\vert \eta \vert < 2.4$.

In order to get images out of the jets we now consider the calorimeter
towers and tracks as jet constituents in the $(\eta, \phi)$ plane. As
in the construction of top taggers \cite{Kasieczka_2017} we will not
use the energy $E$ of the calorimeter towers directly but rather opt
for the transverse energy $E_T = E/\cosh\eta$. The relevant features
are more readily learned by the classifier if we normalize the
coordinates. First, we calculate the $E_T$ weighted center of the
calorimeter towers via
\begin{align} \label{center}
    \overline{\eta}= \frac{\sum_iE_{Ti}\eta_i}{\sum_i E_{Ti}}\,, \\
    \overline{\phi}= \frac{\sum_iE_{Ti}\phi_i}{\sum_i E_{Ti}}\,;
\end{align}
here the sums run over all constituents of a given fat jet. We then
shift the coordinates $\eta_i \rightarrow \eta_i -\overline{\eta}$ and
$\phi_i \rightarrow \phi_i -\overline{\phi}$ so that the image is
centered on the origin.  Next, we rotate the coordinate system around
the origin so that the calorimeter tower with the highest $E_T$ points
vertically from the origin. We use the last degree of freedom to
make sure that the calorimeter tower with the second highest $E_T$
lies in the right half of the coordinate, by flipping along the
vertical axis if necessary.

Next, we pixelate the coordinates to a $0.04\times0.04$ grid. The
brightness/intensity of each pixel is given as the measured $E_T$. We
use three channels, corresponding to $E_T$ in the electromagnetic
calorimeter (ECAL), hadron calorimeter (HCAL), and $p_T$ of the
tracks, analogous to three color channels in classical
images.\footnote{In principle, the tracks have a much higher
  resolution compared to the calorimeter towers and could thus be
  pixelated into a finer grid. However, we do not expect these very
  fine details to improve the discrimination between signal and
  background. We therefore use the same grid spacing for all three
  channels.} We divide each pixel by the maximal value found in this
image, so that each intensity is between $0$ and $1$. This makes
learning more efficient. It also removes information about the $p_T$
and mass of the jet which are powerful discriminators. We partly
remedy this by giving the classifier the mass of the fat jet as
another input; this will be explained in more detail in a following
chapter. We also note that we will later introduce additional
high-level features to our final classifier, which will reintroduce
information about the overall $E_T$ scale of the event.

\begin{figure*}
   \centering
   \includegraphics[width=\textwidth]{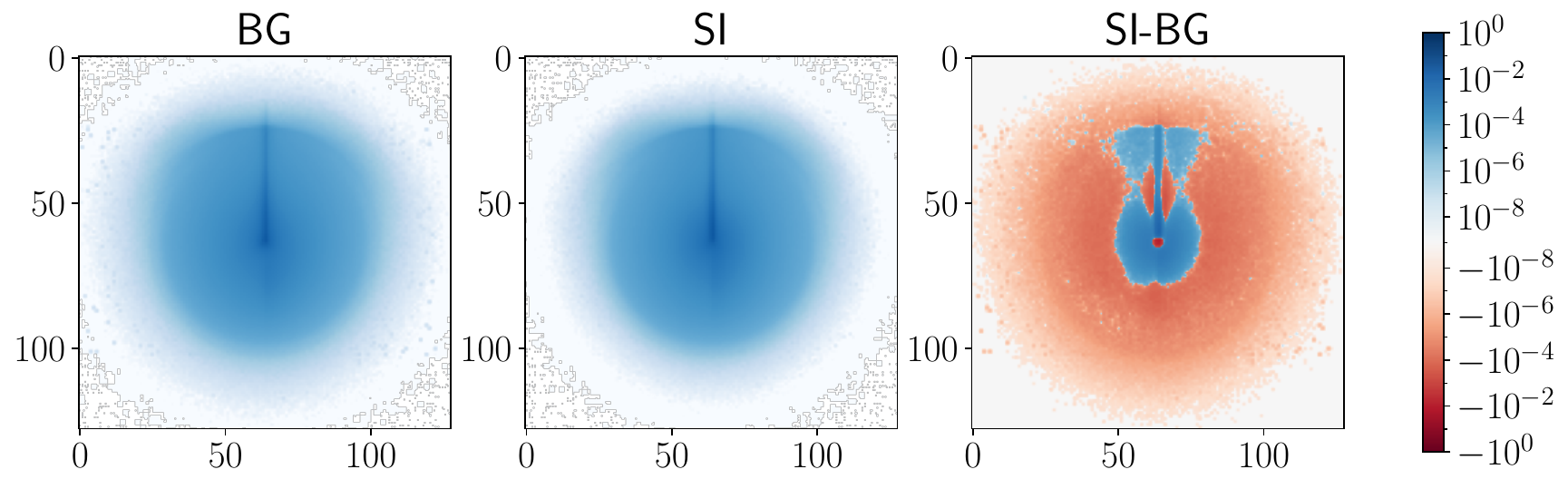}
   \caption{Signal and background AK14 jet image averaged over the
     entire training data set. All three channels are aggregated by
     summation. The rightmost plot shows the difference between the
     average signal and the average background jet image. Signal
     events are more concentrated at the origin while background jets
     are more spread out.}
    \label{fig:jetimage}
  \end{figure*}

In the last preprocessing step we crop the image to a square centered
around the origin with side lengths chosen as 64 pixels for AK08 and
AK10 jets and 128 pixels for AK14 jets. This size is chosen to contain
most of the constituents while also being a power of two which aids in
the application of the computer vision techniques. The resulting
images after all preprocessing steps, averaged over the entire event
sample, are shown in figure~\ref{fig:jetimage}. These average images
look quite similar for signal and background, at least to the human
eye; however, taking the difference between the average images does
reveal some differences.

Moreover, there is more information available to the computer vision
techniques than can be displayed in the figure. For instance, the
number of non-zero pixel values is useful for classification. On
average the signal images contain more non-zero pixels than the the
background images do. Just cutting on this quantity allows, for one
set of parameters, to reach an accuracy of 74\% when applied to a
sample containing equal number of signal and background events. Of
course, our final classifier should perform much better than
this.\footnote{We note in passing that this multiplicity does contain
  some information on the hardness of the event, since it correlates
  positively with both the mass and the transverse momentum of the fat
  jet. Hence the normalization step described above does not
  completely remove the information on these quantities. However,
  these dependencies are only logarithmic, and subject to large
  event-by-event fluctuations. Explicitly adding the jet mass as input
  variable can therefore still be expected to aid in the
  classification task.}

\section{Architectures}
\label{sec:architectures}

In this section we describe how we process a single fat jet, the goal
being to distinguish jets due to the three-body decay of neutralinos
from SM background (``LSP tagging''). At the core is one of three
architectures adapted from computer vision, and described in more
detail in the following three subsections. In all three cases, the
output of this architecture is concatenated with the measured jet mass
and fed into the same multilayer perceptron classification network. It
is built from a dense layer with 256 neurons followed by another dense
layer with 128 neurons which connect to two output neurons. Between
all three layers, the ReLU activation function is used. The two
neurons of the last layer are passed into the softmax activation
function, such that the output can be interpreted as the predicted
probability of the image belonging to either the signal or the
background; since these probabilities should add to $1$, the values of
the two output neurons are almost completely correlated. In the
following, we denote this by \textit{MLP Head}. 

All three architectures use convolution layers. Such a layer convolutes
$c_o$ sets of learnable weights, the kernels, with dimensions
$c_i \times k_h\times k_w$ across the height and width of the image
where $k_h$ and $k_w$ are the height and width of the weights and
$c_i$ is the number of input channels. Recall that we start with three
input channels, akin to the three colors commonly used in vision:
$E_T$ measured in the ECAL, $E_T$ measured in the HCAL, and $p_T$ of
tracks. The information of the three ``colors'' is therefore merged
already in the first convolution layer of a given architecture. The
results of each convolution are stacked such that the output has $c_o$
channels. All architectures are built and trained within the pytorch
\cite{paszke2019pytorch} deep learning library.

\subsection{CNN}
\label{sec:cnn}

\begin{figure}[h]
 \centering
 \includegraphics[width=\linewidth]{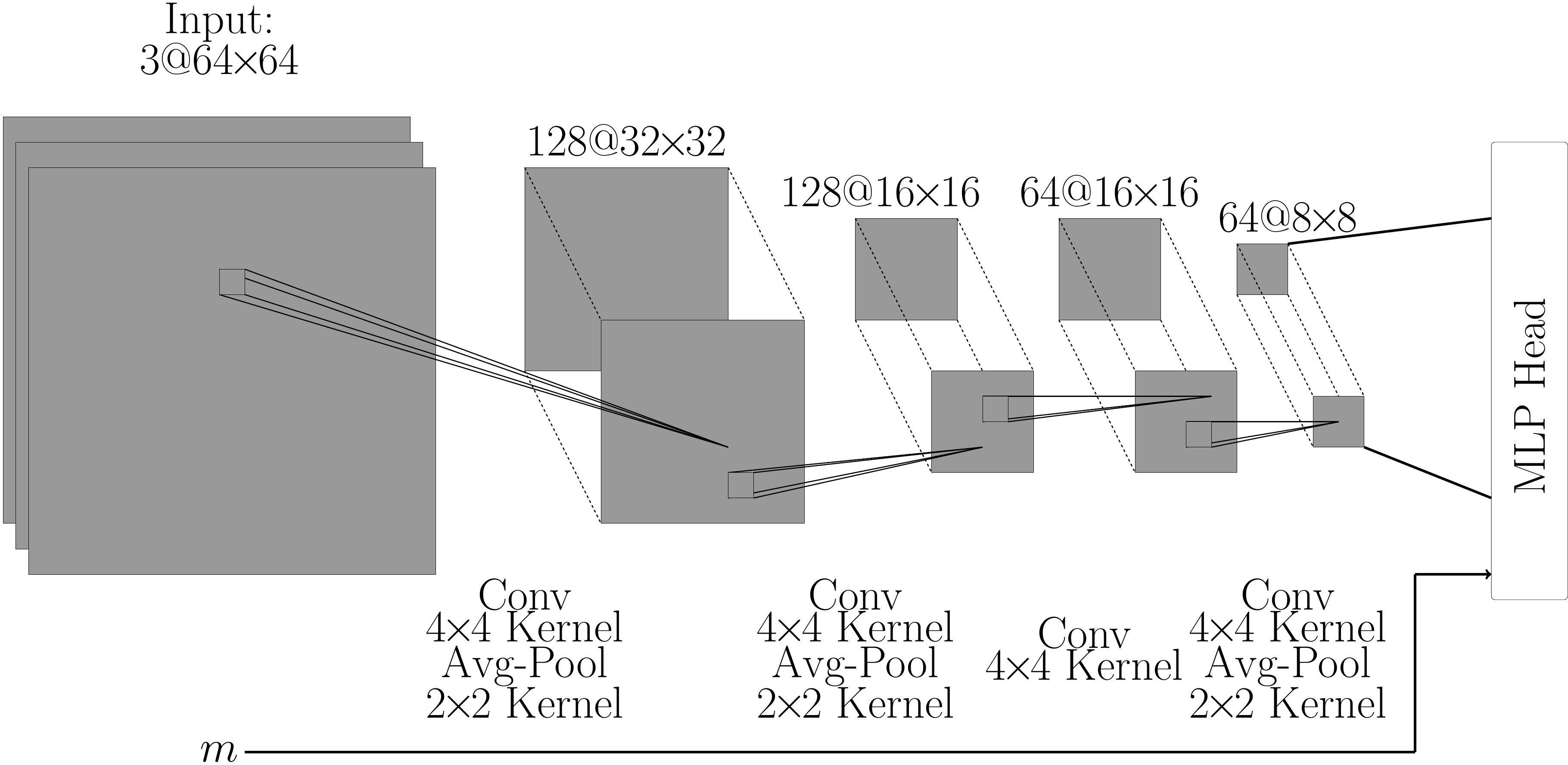}
 \caption{Architecture of the CNN for AK08 and AK10 jets. }
 \label{fig:arch_cnn}
\end{figure}

The first architecture is a (comparatively) simple convolutional
neural net (CNN). We follow loosely an existing model used for
top tagging \cite{Macaluso_2018}. The first layers of the CNN are two
blocks, each containing a convolutional layer with 128 kernels of size
$4\times4$, stride 1, zero padding to keep the image dimensions, ReLU
activation function and average pooling with kernel size 2 and stride
2. This halves the spatial image dimensions. Next, we apply the same
block with only 64 kernels and without pooling. The last convolution
block contains again 64 kernels but this time with the pooling
operation. For AK08 and AK10 jet images, this network produces outputs
of shape $64\times8\times8$; here $8 \times 8$ refers to the size of
one image after three convolutions, and we use $64$ different
``filters'' (i.e. sets of weights in the convolution) for each image. In
order to put AK14 images of size $128\times128$ on the same footing we
repeat the last block one more time. The output of shape
$64\times8\times8$ is then flattened into 4096 features and fed into
the MLP head.

The full architecture for AK08 and AK10 jets is shown in figure
\ref{fig:arch_cnn}. As already noted, this kind of architecture is
already being used for similar tasks; it serves as our baseline,
against which we compare the more advanced architectures described in
the following two subsections.

\subsection{CoAtNet}

Since their inception in the context of natural language processing
\cite{vaswani2023attention}, transformer models have been shown to
also be applicable for computer vision tasks. Because the attention
mechanism inside the transformer is computationally expensive it is
often impractical to apply it directly to the entire input image. In
order to counteract this, the vision transformer (ViT) model
\cite{dosovitskiy2021image} splits the input image into manageable
patches. They are then flattened, linearly projected, equipped with a position embedding and fed into the transformer
encoder. Since both translation equivariance and locality are
explicitly broken in this approach one cannot expect to outperform
CNNs that are known to leverage both of these features, and therefore
generalize well onto the unseen test set. In our tests, the vanilla
ViT indeed performs poorly, so we will not pursue it further.

However, it is possible to construct a model that combines the global
receptive field of view from self-attention with the aforementioned
advantages of CNNs. This is the aim of CoAtNet \cite{dai2021coatnet}
(the name derives from the combination of depth-wise
\textbf{Co}nvolution and self-\textbf{At}tention). Let us now briefly
describe how CoAtNet is built; we refer the interested reader to the
original publication for a more detailed description.

The model is constructed in five stages. The first stage consists of
three convolution layers which $3\times 3$ kernels, where the first
has a stride of $2$. This halves the spatial resolution of the input
image. This is followed by two stages of three MBConv blocks
\cite{sandler2019mobilenetv2} which are computationally cheaper while
maintaining most of the performance of full convolutional layers. In
both stages, the first layers perform downsampling again with a stride
size of $2$. By now the width and height of the input image are shrunk
by a factor of $2^3$ so global attention is feasible even for the
large AK14 jet. Thus the last two stages consist of five and two
transformer blocks respectively. In each transformer 2D relative
attention is used which adds a learnable weight that depends only on
the relative 2D position. These steps were performed using the
publicly available
code\footnote{\url{https://github.com/chinhsuanwu/coatnet-pytorch}}. Finally,
we average pool the outputs and feed the 768 features into the
classification head.

\subsection{MaxViT}

The third architecture we will use is the Multi-Axis Vision
Transformer (MaxViT) \cite{tu2022maxvit}. This approach circumvents
the shortcomings of the vanilla ViT differently. Instead of splitting
the image in patches and only applying attention to these flattened local patches,
the goal will be to apply it locally and globally consecutively
by decomposing the image with two strategies into smaller bits. First,
the image is separated into equal-sized, non-overlapping blocks along
the spatial dimensions. Self-attention is applied within each block
using only the local information. In order to leverage global
information the next strategy groups every pixel that is reached by a
step of fixed size into the object to which self-attention is
applied. (For a nice visualization we refer to Fig.~3 of
\cite{tu2022maxvit}). This approach essentially uses high-resolution
local information before using low-resolution global information.
Both attention mechanisms are applied after a MBConv-block forming the
so-called MaxViT-block.

For our application, we chose the publicly available implementation
that is shipped with
pytorch\footnote{\url{https://github.com/pytorch/vision/blob/main/torchvision/models/maxvit.py}}
with the only change being the replacement of the MLP head that takes
the 256 output features. Concretely, the first stage consists of two
convolutional layers with 64 $3\times3$ kernels each, where the first
has stride $2$, reducing the spatial dimensions by half. This is
followed by three stages with two MaxViT blocks each. The convolution
is strided with size 2 for the first block of each stage. The
partitions are of size $4\times4$ each. The MaxViT stages have 64,128
and 256 channels respectively. Self-attention uses 32-dimensional
heads.

\section{Data set creation}
\label{sec:dataset_creation}

In this section we describe how the data set used to train the LSP
taggers is defined. In most events, there is more than one fat jet
that passes the selection criteria. Since we investigate pair
production, not all the information useful for event classification
can be expected to be contained in the hardest jet. It is therefore
expected to be useful to combine information from more than one jet
into the analysis. Wide jets that are produced from the
$\tilde\chi$ decays are expected to be hard because of the large
stop mass. Therefore the two jets with the largest $p_T$ are expected
to be signal enriched. The preselection requirements imply that one
top quark from stop decay generally will decay semileptonically.
However, the second top quark might decay fully hadronically,
resulting in a third wide jet with large $p_T$. Since the top quarks
and the LSPs have very similar $p_T$ distributions, the third
largest-$p_T$ jet may also well be from an LSP.

In order to design taggers that perform well on all three leading
jets, and hence for a wide range of $p_T$, we therefore include
samples of all three leading fat jets in our training data set, in the
ratios that the respective number of jets are present in the full
events. To this end, we add up to three fat jets present in an event
as images to the data set. Of course, an event may also contain only
one or two such jets; in fact, this is generally the case for
background events. We generate as many events as required to reach the
desired size of the training set for each jet size. 

\section{Training the LSP taggers}
\label{sec:train_base}

We start by verifying that the different \textsc{pythia} tunes that we
adopted do not significantly influence our results. To this end, we
train the CNN model described in sec.~\ref{sec:cnn} to differentiate
not between signal and background samples but between background
events generated with the CP2 tune and background samples generated
with the CP5 tune. We combine \numprint{1000000} jet images generated
with each tune into a data set and split it equally between training
and test sets for each jet size. The initial learning rate $\eta_L$ is
chosen as $5\cdot 10^{-4}$. This value worked best in tests of the LSP
taggers. At the end of each epoch the learning rate is lowered by a
factor of $0.7$ and the entire training data set is shuffled. The batch
size is $64$. We minimize the averaged binary cross-entropy loss
\begin{equation}
    l = -\frac{1}{N}\sum^N_i y_i\ln x_i + (1-y_i)\ln(1-x_i)\,,
    \label{eq:loss}
\end{equation}
where the index $i$ runs through all $N=64$ images in the batch, $y_i$ is
the true label and $x_i$ is the predicted label. Adam
\cite{kingma2017adam} is chosen as the optimizer. All taggers are
trained for a total of $15$ epochs.

The minimum validation losses for AK08, AK10 and AK14 jets are found
to be $0.6917$, $0.6918$ and $0.6920$, respectively. When the
classifier is tasked to assign the label $0$ to the first class
(e.g. CP2 tune) and the label $1$ to the second class (CP5 tune) and
the classifier is perfectly confused (i.e. unable to distinguish
between the classes), it will assign labels close to $0.5$ regardless
of the true class. The binary cross entropy per image is then
$\ln(2) = 0.6931$. Evidently our observed losses are only very
slightly below the value expected for a classifier that learns
nothing. We therefore conclude that the difference in \textsc{pythia}
tunes can be neglected in the following.

We now turn to the actual training of the taggers to select LSP-like
fat jets. Signal samples are generated for $\tilde\chi$ masses
between \SI{100}{\giga \eV} and \SI{500}{\giga \eV} in \SI{10}{\giga
  \eV} steps, and for stop masses between \SI{700}{\giga \eV} and
\SI{1200}{\giga \eV} in \SI{25}{\giga \eV} steps. For each combination
of stop and neutralino mass, we take $4750$ sample images from
$\tilde t_1 \tilde t_1^*$ signal events. In order to generate an
almost pure signal sample for training, we only include images of jets
that are within $\Delta R<0.5$ of a parton level
$\tilde\chi$. Since we want the LSP tagger to work for all
combinations of $m_{\tilde t_1}$ and $m_{\tilde \chi}$, we combine
all $41 \times 21 \times 4750 = \numprint{4089750}$ images into a
single training set. We take the same number of background images,
\numprint{4089750}, from $t \bar t + $jets events.

Finally, we split the \numprint{8179500} images into
\numprint{5725650} images for training and \numprint{2453850} images
for validation. After training the model state at the epoch with the
lowest validation loss is selected to define the tagger.

\section{Results for neutralino taggers}
\label{sec:res_base_classifiers}

In order to compare the performance of our classifiers we neglect any
systematic uncertainties and define the signal significance $Z$ as
\begin{equation} \label{eq:Z}
  Z=\frac {S} {\sqrt{B}} = \frac {\epsilon_S} {\sqrt{\epsilon_B}}
  \cdot \frac {\sigma_S} {\sqrt{\sigma_B}} \sqrt{\mathcal{L_{\rm int}}}\,,
\end{equation}
where $S$ and $B$ are the number of signal and background samples
passing a cut (e.g. on the value of an output neuron of the MLP),
$\epsilon_{S/B}$ is the selection efficiency of this cut,
$\sigma_{S/B}$ is the fiducial production cross section and
$\mathcal{L_{\rm int}}$ is the integrated luminosity of the data set
considered. Instead of comparing signal significances directly, we
compare the significance improvement $\epsilon_S/\sqrt{\epsilon_B}$,
which captures the gain due to the sophisticated event classifiers,
and is independent of the assumed luminosity.

\begin{figure}
\centering
\includegraphics[width=\linewidth]{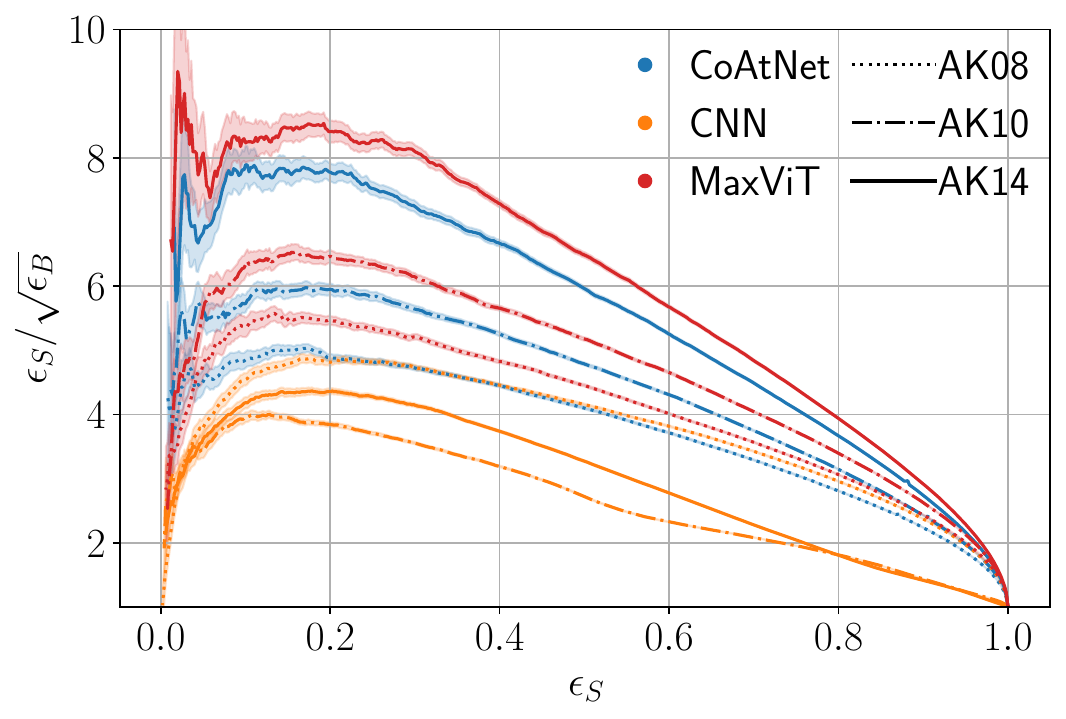}
\caption{Significance improvement curves for all three neutralino
  taggers for all single jet samples in the test data set. The shaded
  regions show one bootstrapped standard deviation.}
\label{fig:baseclassifier_sic}
\end{figure}

Figure \ref{fig:baseclassifier_sic} shows the performance of all
neutralino taggers on the entire test data set, i.e. with all signal
masses present and with the three leading jets mixed as mentioned in
sec.~\ref{sec:train_base}. As a working point for the following
analysis we choose the cut on the MLP output neuron such that
$\epsilon_S=0.3$. Even lower values of $\epsilon_S$ can still increase
$\epsilon_S/\epsilon_B$, but the significance improvement is already
close to the maximum at the chosen point. Moreover, for smaller
$\epsilon_S$ the background efficiency $\epsilon_B$ becomes so small
that the statistical uncertainty on the accepted background becomes
sizable, in spite of the large number of generated background
events.

Both CoAtNet and MaxViT showed superior performance in classical image
classification tasks compared to CNN-based models, as is reported in
the respective original publications. We expect this to carry over to
jet classification. Indeed this is the case here and both models
outperform the classical CNN by up to a factor of $2$ for AK14 jets.
The most performant classifiers are the transformer-based models
trained on the large radius jets. These large jets still contain the
entire narrow jets from small LSP masses while the small jets might
miss important features for larger neutralino masses.  We also observe
that MaxViT performs slightly better than CoAtNet, as is the case in
the original MaxViT publication. Evidently improvements in modern
computer vision translate well to the classification of jet
images. Even the worst transformer-based model (i.e. AK08 CoAtNet)
matches the best CNN. Interestingly, despite the transformer models
showing a clear hierarchy, the larger a jet is the better, this is not
the case for the CNN, which performs best for AK08 jets.

So far we have considered classification of singlet jets. In the next
section, we will show how this can be used for event classification.

\section{Boosted classifiers}
\label{sec:boosted_classifiers}

As previously mentioned our signal model always produces two
neutralinos that subsequently decay into three quarks (plus possible
gluons from final state radiation). It is therefore instructive to
combine multiple jet images into our predictions. To this end we apply
one of our LSP taggers described above on the three leading fat jets
in an event; from now on we drop the merging requirement since it is
not meaningful anymore. The three resulting MLP outputs are used as
inputs for a gradient-boosted decision tree (GBDT) classifier. If an
event contains less than $3$ fat jets with $p_T > 100$ GeV we assign
the label $-1$ for the missing jets. The GBDT is implemented using the
XGBoost \cite{Chen_2016} package. We use $120$ trees with a learning
rate of $0.1$ with other hyperparameters left unchanged from the
default values. In order to train the GBDT and calculate its results
we use $3000$ and $2500$ events, respectively, for each combination of
stop and LSP masses. This corresponds to a total of \numprint{2583000}
signal events for training and \numprint{2152500} signal events for
calculating results. We again generate an equal number of
$t \bar t + $ jets background events.

\begin{figure}
  \centering
  \includegraphics[width=\linewidth]{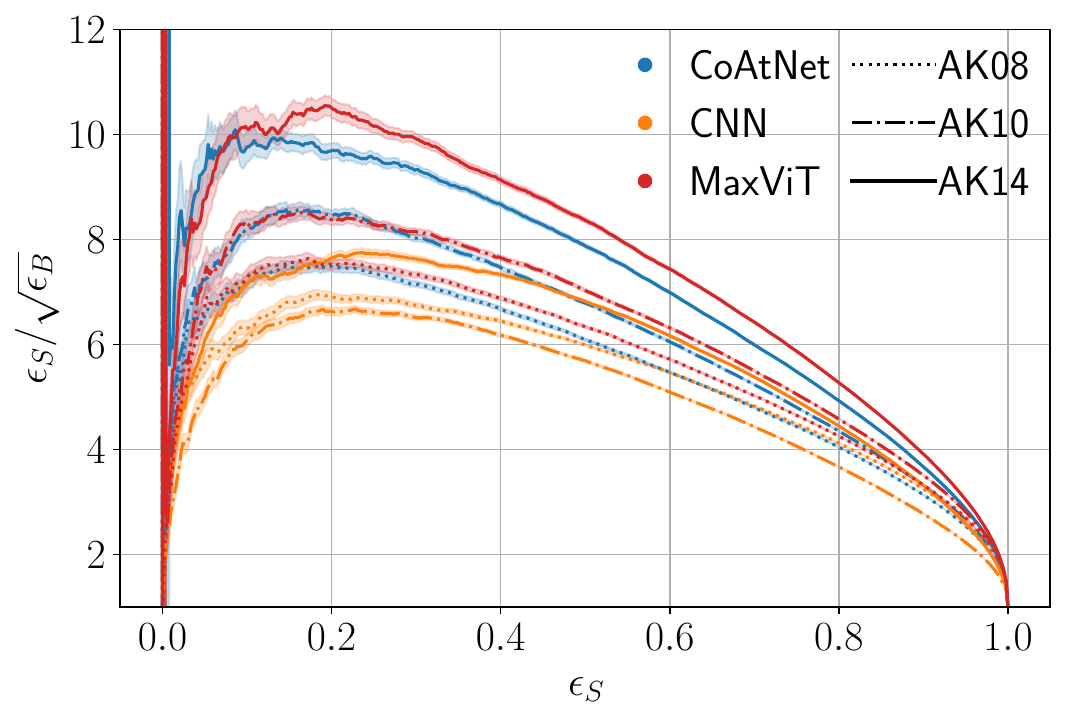}
  \caption{Significance improvement curves for all GBDT classifiers
    built to combine the LSP tagger outputs for the three highest
    $p_T$ jets. The shaded regions are one bootstrapped standard
    deviation.}
\label{fig:baseclassifier_sic_GBDT}
\end{figure}

Figure \ref{fig:baseclassifier_sic_GBDT} shows the significance
improvement after a cut on the signal probability given by the
GBDT. The difference in performance between the two transformer-based
models has shrunk significantly for all jet sizes, especially for AK10
and AK08 jets. Comparing this with figure \ref{fig:baseclassifier_sic}
the gain by combining the three jets is not very large. One has to
keep in mind, that the merging requirement is now dropped. If we
calculate the significance improvement for only the jet with the
highest $p_T$ without requiring it to be close to a (truth-level) LSP,
MaxViT reaches $6.79 \pm 0.05$ at $\epsilon_S=0.3$. Comparing this
with $9.92 \pm 0.12$ for the same base model after combining the LSP
tagger output for the three hardest jets shows an improvement of
almost 50\%, equivalent to doubling the integrated luminosity in
eq.(\ref{eq:Z}). The CNN now also works best with AK14 jets, even
though the AK08 version is still better than the AK10 version,
contrary to the hierarchy of the other models.

Overall the level of improvement between the results of
fig.~\ref{fig:baseclassifier_sic_GBDT}, which use information from up
to three jets per event, and fig.~\ref{fig:baseclassifier_sic} for
single jets, might seem somewhat disappointing. After all, in the
absence of QCD radiation a $\tilde t_1 \tilde t_1^*$ signal event
contains two signal jets plus one fat background jet from the
hadronically decaying top quark, whereas a generic $t \bar t$ event
with one top quark decaying semileptonically contains only a single
background fat jet. In such a situation simply requiring at least one
fat jet to be tagged as signal would increase the signal efficiency
(for $\epsilon_S \gg \epsilon_B$) from $\epsilon_S$ to
$1 - (1-\epsilon_S)^2$ while the background efficiency remains
unchanged. Recall, however, that we require each event to contain at
least seven AK04 jets. This greatly reduces the $t \bar t$ background,
since at least three additional partons need to be emitted for the
event to pass this cut; on the other hand, it also means that
background events frequently contain several fat jets, in which case a
simple single tag requirement would not increase the significance. In
any case, as noted above there is a significant improvement in
performance when information of the three leading fat jets is combined
using a GBDT; of course, the GBDT output is not equivalent to simply
demanding a fixed number of jets in a given event being tagged as
LSP-like.

\begin{figure}
 \centering
\includegraphics[width=\linewidth]{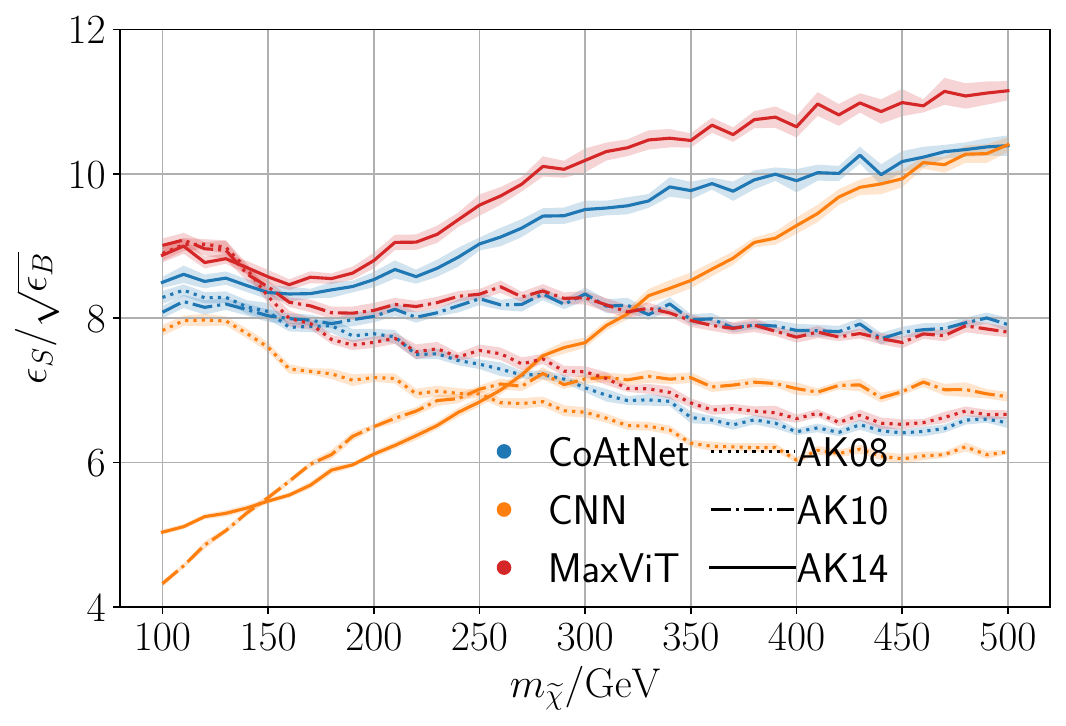}
\caption{Significance improvements depending on the LSP mass for all
  GBDT classifiers built to combine the LSP tagger outputs for the
  three highest $p_T$ jets. The cut on the GBDT output has been set
  such that $\epsilon_S=0.3$ for each given LSP mass. The shaded regions are one
  bootstrapped standard deviation.}
\label{fig:classifiers_mass}
\end{figure}

In Fig.~\ref{fig:classifiers_mass} we show how the performance of the
GBDT depends on the LSP mass. For small masses, the two
transformer-based models perform comparably for all three jet
sizes. Here the decay products are usually contained even in the AK08
jet so all three jet sizes contain the necessary information for our
task. Evidently the transformer networks are able to filter out the
noise from particles not related to LSP decay that are present in the
AK10 and AK14 jets, while the simpler CNN cannot; hence the GBDT using
the CNN applied to AK10 or AK14 jets performs relatively poorly for
small LSP mass. On the other hand, for LSP mass above 200 GeV the GBDT
performs significantly worse when used on the smaller jets, which no
longer contain all particles originating from LSP decay.

We also note that using the CNN applied to AK14 jets performs far
worse than the other models for small LSP mass, but matches the
performance of the CoAtNet-based model for $m_{\tilde \chi}$ between
450 and 500 GeV. This curve also shows the strongest LSP mass
dependence. We will revisit this point later in this chapter.

Finally, while the MaxViT architecture with AK10 and AK14 jets again
shows the best overall performance, the resulting
$\epsilon_S/\sqrt{\epsilon_B}$ shows a shallow minimum at
$m_{\tilde \chi} \simeq m_t$. For a given $p_T$ fat jets originating
from LSP and top decay will then have similar overall features, and
the additional information about the jet mass will not help at all;
moreover, recall that in our scenario the LSP decay products contain
exactly one $b$-quark, just like nearly all jets from top decay.
Nevertheless the model performs quite well even in this difficult mass
region. Presumably it exploits the fact that top decays into three
quarks proceed via two 2-body decays with a color singlet on-shell $W$
boson in the intermediate state, whereas the LSP decays via the
exchange of a (far) off-shell squark.
 
\begin{table}
 \begin{center}
  \begin{tabular}{ |c c c| }
   \hline
   Model & AK14 & Combined jets\\
   \hline
   CoAtNet&$9.32 \pm 0.10$&$9.63 \pm 0.10$\\
   MaxViT&$9.91 \pm 0.11$&$10.09 \pm 0.12$\\
   CNN&$7.62 \pm 0.06$ & $9.16 \pm 0.09$\\
   \hline
  \end{tabular}
  \caption{Significance improvement, $\epsilon_S / \sqrt{\epsilon_B}$,
    for $\epsilon_S=0.3$ when only using AK14 jets (second column),
    and when combining the LSP tagger outputs on AK08, AK10 and AK14
    jets using a larger GBDT (third column). The uncertainties are
    bootstrapped standard deviations.}
        \label{tab:combine_jetdefs}
    \end{center} 
\end{table} 

At this point we still have nine predictions for each event (the output
of three architectures applied to AK08, AK10 and AK14 jets). Of
course, these nine numbers are highly correlated. Nevertheless a further
improvement of the performance might be possible by either combining
results from different jet definitions within a given architecture, or
vice versa. Comparing these results might also allow us to infer in
which aspect a single model has room for improvements that might be
gained by another architecture.

We start by combining LSP tagger outputs for different jet sizes.  We
show the results in table \ref{tab:combine_jetdefs} and compare the
performance to that of the best single jet definition, which is
achieved for AK14 jets as we saw in
fig.~\ref{fig:baseclassifier_sic_GBDT}. Evidently the improvement is
barely statistically significant for the two transformer-based
models. These models extract most of the useful information from the
images of the large AK14 jets, even when there is a lot of clutter
present. The improvement is larger for the CNN-based classifier, which
however still performs somewhat worse than the other models. It seems
to benefit from the multiple jet definitions intended to extract
high-level features such as the mass in classical applications. In
particular, the combination allows to compensate the degraded
performance when using the large jets for LSP mass below
$\SI{250}{\giga \eV}$ by information from the AK08 jets which is more
useful in this parameter region, as we saw in
fig.~\ref{fig:classifiers_mass}.

\begin{table}
 \begin{center}
  \begin{tabular}{ |c c c| }
   \hline
   Jet &  Combined & Best single model\\
   \hline
   AK08 & $7.53 \pm 0.06$ & $7.34 \pm 0.06$ (MaxViT)\\
   AK10 & $8.94 \pm 0.09$ & $8.15 \pm 0.07$ (MaxViT)\\
   AK14 & $10.51 \pm 0.11$ & $9.91 \pm 0.11$ (MaxViT)\\
   \hline
  \end{tabular}
  \caption{Significance improvement with $\epsilon_S=0.3$ when feeding
    the outputs of all three LSP taggers simultaneously to the GBDT,
    keeping the jet definition fixed. For comparison the third column
    shows the significance improvement for the MaxViT-based model,
    which performs best for all three jet sizes. The uncertainties are
    bootstrapped standard deviations.}
  \label{tab:combine_models}
 \end{center} 
\end{table}  

Next, we combine the outputs of different LSP taggers into a single
GBDT, for fixed jet definition. The results are shown in
table~\ref{tab:combine_models}. This time the combination leads to a
slight but significant improvement over the best single model (the one
based on MaxViT). This shows that even though the transformer-based
models perform almost equally well for the AK14 jets while the CNN is
noticeably weaker, each model misses complementary information that
the GBDT can combine into a stronger classifier.

\begin{figure}
 \centering
  \includegraphics[width=\linewidth]{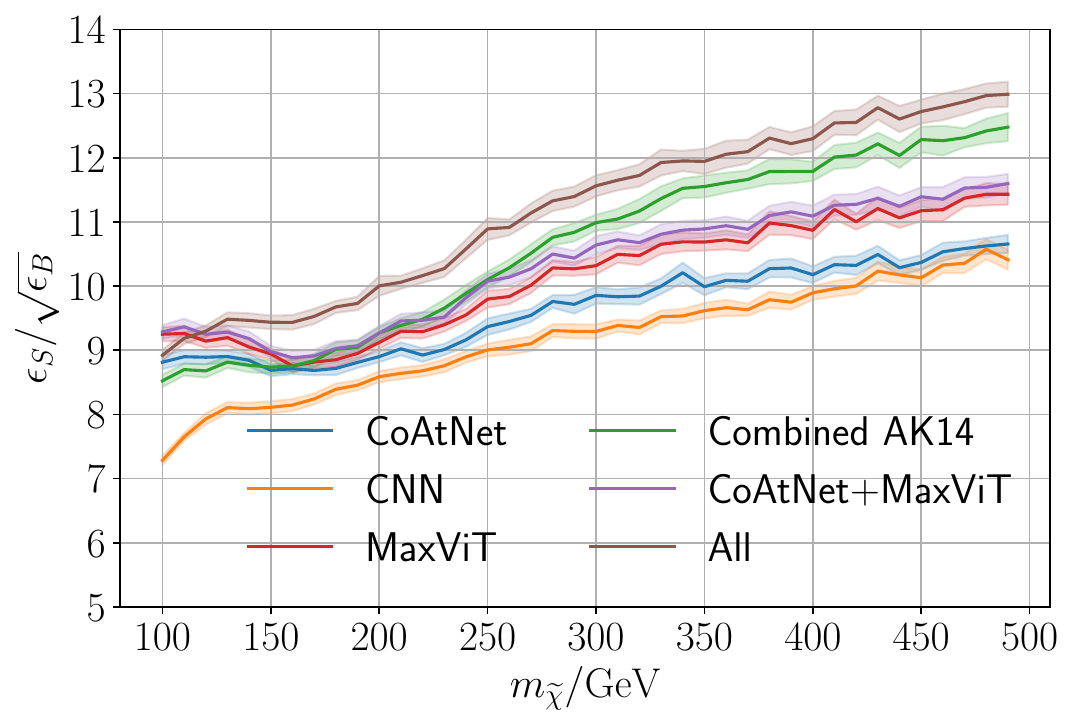}
  \caption{Significance improvement as function of the LSP mass for
    GBDT classifiers built by combining the output of different LSP
    taggers, with $\epsilon_S=0.3$ in each case. The shaded regions
    are one bootstrapped standard deviation. The curves labeled
    CoAtNet (blue), CNN (orange) and MaxViT (red) show the performance
    of GBDTs built from combining the jet sizes for the given model,
    as in the third column of table~\ref{tab:combine_jetdefs}. The
    green curve is for the GBDT that uses the outputs of all LSP
    taggers, but only for the AK14 jets, as in the third row of
    table~\ref{tab:combine_models}. The purple line results from
    combining both transformer-based LSP taggers for all jet sizes,
    while the brown line is for a GBDT that combines all LSP taggers
    and all jet sizes. }
 \label{fig:classifiers_combined_all_mass}
\end{figure}

In figure \ref{fig:classifiers_combined_all_mass} we show how the
performance of various strategies to combine LSP taggers varies with
the neutralino mass. Combining all transformer-based predictions into
a single GBDT does not show any significant improvement over the
performance of the MaxViT-based tagger. This indicates that these
models use the same features of the jet images and do not find
complementary information. The combination of all CNN predictions is
comparable to the weaker transformer-based model, CoAtNet, for LSP
mass above $\SI{200}{\giga \eV}$, while MaxViT is still more sensitive
for all LSP masses.

Because our LSP taggers generally perform best on AK14 jets we also
show the combination of all three architectures using only AK14 jets,
as in the last row of table~\ref{tab:combine_models}. Comparing to
fig.~\ref{fig:classifiers_mass} we see that for LSP mass below
$\sim \SI{160}{\giga \eV}$ this combination does not further improve
on the MaxViT-based model applied to AK14 jets. Between
$\sim \SI{160}{\giga \eV}$ and $\sim \SI{300}{\giga \eV}$ the
performance closely follows that of the two combined transformer
models shown in purple. Since we already showed that one does not gain
much combining the CoAtNet and MaxViT models this shows that the CNN
does not yield useful information in this region of parameter space,
either.

However, as we saw in fig.~\ref{fig:classifiers_mass} the CNN-based
model applied to AK14 jets improves more with increasing LSP mass than
the transformer-based models do, even matching CoAtNet at
$\SI{500}{\giga \eV}$. The combination profits from this fact and
outperforms above $\SI{300}{\giga \eV}$ the GBDTs using only input from
the transformer-based LSP taggers. This shows that the CNN learns
something about the sample that the other models miss.

Finally, we show the result of a GBDT that is trained on the LSP tagger
outputs of all three models and all three jet sizes, and thus has $27$
inputs in total for each event. Compared to the AK14 only case this
does benefit from the inclusion of smaller jets, in particular at
smaller LSP masses where the AK08 and AK10 jets already capture
most LSP decay products. For larger LSP masses the performance is only
slightly better than that of the AK14-only case.
 
These various comparisons show that for the given signal process, the
largest improvement in significance $\epsilon_S/\sqrt{\epsilon_B}$ is
achieved by the transformer-based models applied to AK14 jets. Both
models capture details of the jet images that the CNN misses. Nevertheless
also feeding the output of the CNN-based LSP tagger into a larger GBDT
leads to a further slight improvement of the performance. This indicates
that one might be able to find new architectures that perform even better
than MaxViT.

\section{Adding high-level features}
\label{sec:add_high_level_features}

The cuts discussed in section \ref{sec:datagen_preselection} are only
preselections. They ensure that the event passes the single lepton
trigger and contains at least one fat jet to which the LSP tagger can
be applied. They also reduce the background, but even after including
information from the LSP tagger these cuts are not likely to yield the
optimal distinction between signal and background. A full event has
additional features that allow to define additional, potentially
useful cuts, even if they may show some correlation with the output of
the LSP tagger.

In particular, so far the only dimensionful quantities we used in the
construction of our classifier are the masses of the hardest three fat
jets, which we use as input of the LSP tagger. We therefore now
introduce as additional input variables for the final GBDT the sum of
the masses of all AK14 jets \cite{Hook_2012},
\begin{equation}
    M_J=\sum_{\rm AK14} m\,,
\end{equation}
and the total missing transverse momentum $p_T^{\rm miss}$. In
addition, we use the total number $N_j$ of all AK04 jets as well as
the scalar sum $H_T$ of their transverse momenta.

Moreover, information about the angular separation of the jets might
be helpful.  Inspired by ref. \cite{PhysRevD.104.032006} we capture
this information via the Fox-Wolfram moments \cite{Bernaciak_2013}
$H_l$, defined by
\begin{equation} \label{eq:Hl}
 H_l = \sum_{i,j=1}\frac{p_{Ti}p_{Tj}}{(\sum_k p_{Tk})^2}P_l(\cos\Omega_{ij})\,;
\end{equation}
here $i,\,j,\,k$ run over all AK04 jets in the event, $p_{Ti}$ is the
$p_T$ of the $i$.th jet, $P_l$ is the Legendre polynomial and
\begin{equation} \label{eq:Omega}
  \cos\Omega_{ij} = \cos\theta_i \cos\theta_j
  + \sin\theta_i \sin\theta_j \cos(\phi_i-\phi_j)
\end{equation}
is the cosine of the opening angle between the jets $i$ and $j$.

We combine these features into two sets. First the small set
DS1$=[p_T^{\rm miss}, H_T, M_J, N_j]$ which includes the most commonly
used features for new physics searches in hadronic final states. In
addition, we also consider a slightly larger set DS2, which also
includes the second to sixth Fox-Wolfram moments. We combine these
features with the output of the LSP tagger based on MaxViT applied to
AK14 jets (i.e. the most performant single model) and derive
predictions with a similar GBDT as before.

\begin{figure}
 \centering
 \includegraphics[width=\linewidth]{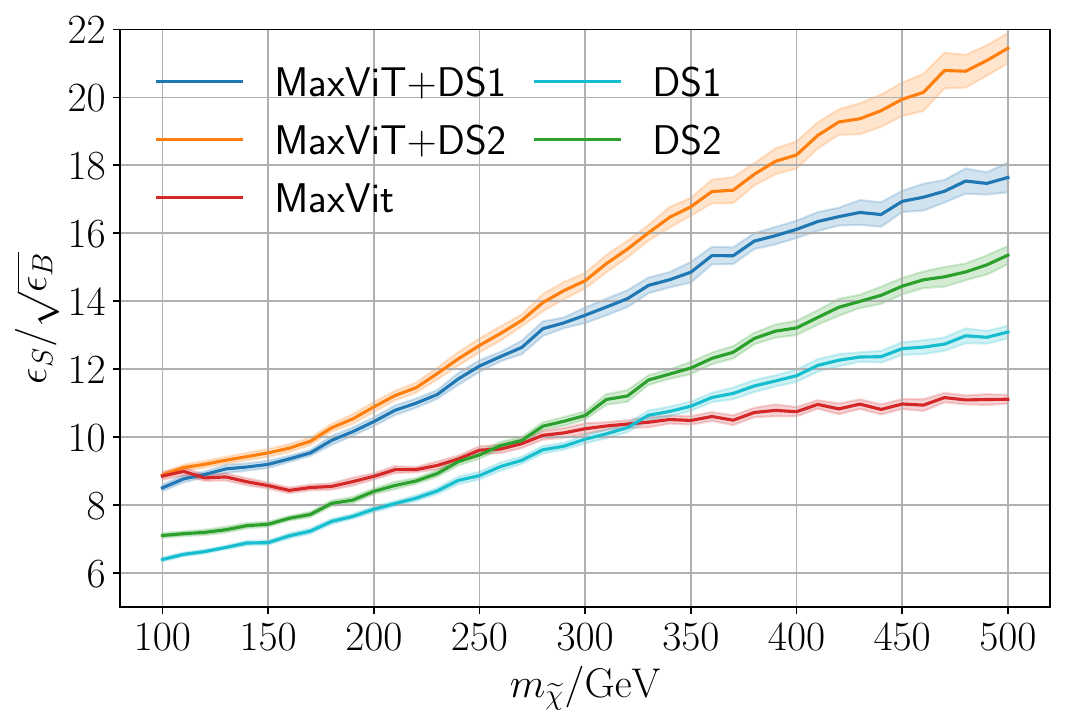}
 \caption{Significance improvements as function of the LSP mass for
   various GBDT classifiers. In all cases the cut on the GBDT output
   has been set such that $\epsilon_S=0.3$ for each given LSP
   mass. The upper two curves show results from classifiers that
   combine the output of the MaxViT-based LSP tagger applied to AK14
   jets with additional kinematical features. The feature set DS1
   contains $[p_T^{\rm miss}, H_T, M_J, N_j]$ while DS2 contains, in
   addition, the second to sixth Fox-Wolfram moments. For comparison,
   the red curve results when using only LSP tagger information, as in
   fig.~\ref{fig:classifiers_mass}, while the lower blue and green
   curves are for GBDTs that only use kinematical information. The
   shaded regions are one bootstrapped standard deviation.}
    \label{fig:classifiers_combined_all_mass_additional_data}
\end{figure}

Results are shown in
fig.~\ref{fig:classifiers_combined_all_mass_additional_data}. We see
that even GBDT classifiers that only use the kinematic information of
sets DS1 or DS2 are quite capable of separating signal from
background, especially for larger LSP masses; this reconfirms the
usefulness of these variables for new physics searches at the LHC. In
fact, for LSP mass above $300$ GeV these classifiers even outperform
the GBDT that only uses information from the MaxViT-based LSP tagger.
On the other hand, except for $m_{\tilde \chi} = 100$ GeV adding
kinematic information to the output of the LSP tagger clearly improves
the performance of the event classifier; the Fox-Wolfram moments prove
useful for LSP mass above $250$ GeV or so.

Conversely, adding information from the LSP tagger to the purely
kinematic variables raises the significance improvement by an amount
which is nearly independent of the LSP mass. We expect the gain of
performance to be even larger when compared to a classical selection
based purely on kinematical cuts.

\section{Application at 137 \texorpdfstring{fb$^{-1}$}{1/fb}}
\label{sec:application_137}

We are now ready to discuss how the different classifiers fare, in
terms of the reach in stop mass for exclusion or discovery. Here we
set the integrated luminosity to
$\mathcal{L}_\text{int}=\SI{137}{\per \femto \barn}$, as in the
original CMS publication \cite{PhysRevD.104.032006}. For simplicity we
ignore the systematic uncertainty on the signal, as well as the
uncertainty from the finite size of our Monte Carlo samples. The
former is much less important than the systematic error on the
background estimate, and the latter should be much smaller than the
statistical uncertainty due to the finite integrated luminosity. The
$t\bar{t}$ background is normalized to the next to leading order
production cross-section \cite{Alwall_2014}. The simulated stop pair
samples are normalized to NLO + NLL accuracy
\cite{Borschensky_2014}. This corresponds to \numprint{273084}
background events and a stop mass dependent number of signal events.
We calculate exclusion limits from the expected exclusion significance
\cite{Kumar_2015}:
\begin{align} \label{eq:Zex}
%  Z_{\rm excl} = &\left[ 2\left( S - B \ln\left( \frac{B+S+x}{2b}\right)
%    -  \frac {B^2} {\Delta_B^2} \ln \left(\frac{B-S+x}{2B}\right)
  Z_{\rm excl} = &\left[ 2S - 2B \ln\left( \frac{B+S+x}{2B}\right)
    -  \frac {2B^2} {\Delta_B^2} \ln \left(\frac{B-S+x}{2B}\right)
   \right. \nonumber \\ & \left.
  -(B+S-x) \frac {B+\Delta_B^2} {\Delta_B^2}  \right]^{1/2}\,,
\end{align}
where $B$ and $S$ are the expected number of background and signal
events, $\Delta_B$ is the absolute systematic uncertainty on the
background, and
\begin{equation} \label{eq:x}
 x= \sqrt{ (S+B)^2- \frac {4SB\Delta_B^2} {B+\Delta_B^2}} \,.
\end{equation}
We chose $\Delta_B = 0.06 B$, as described below. $Z_{\rm excl}$ is the
expected number of standard deviations with which the predicted signal
$S$ can be excluded if the background-only hypothesis, described by
the background $B$, is correct; note that
$Z_{\rm excl} \rightarrow S/\sqrt{B + \Delta_B^2}$ if $B \gg S$. This
quantity is computed for every combination of stop and LSP masses
introduced in sec.~\ref{sec:datagen_preselection}, using four
different event classifiers.

\begin{figure*}
 \centering
 \includegraphics[width=0.48\linewidth]{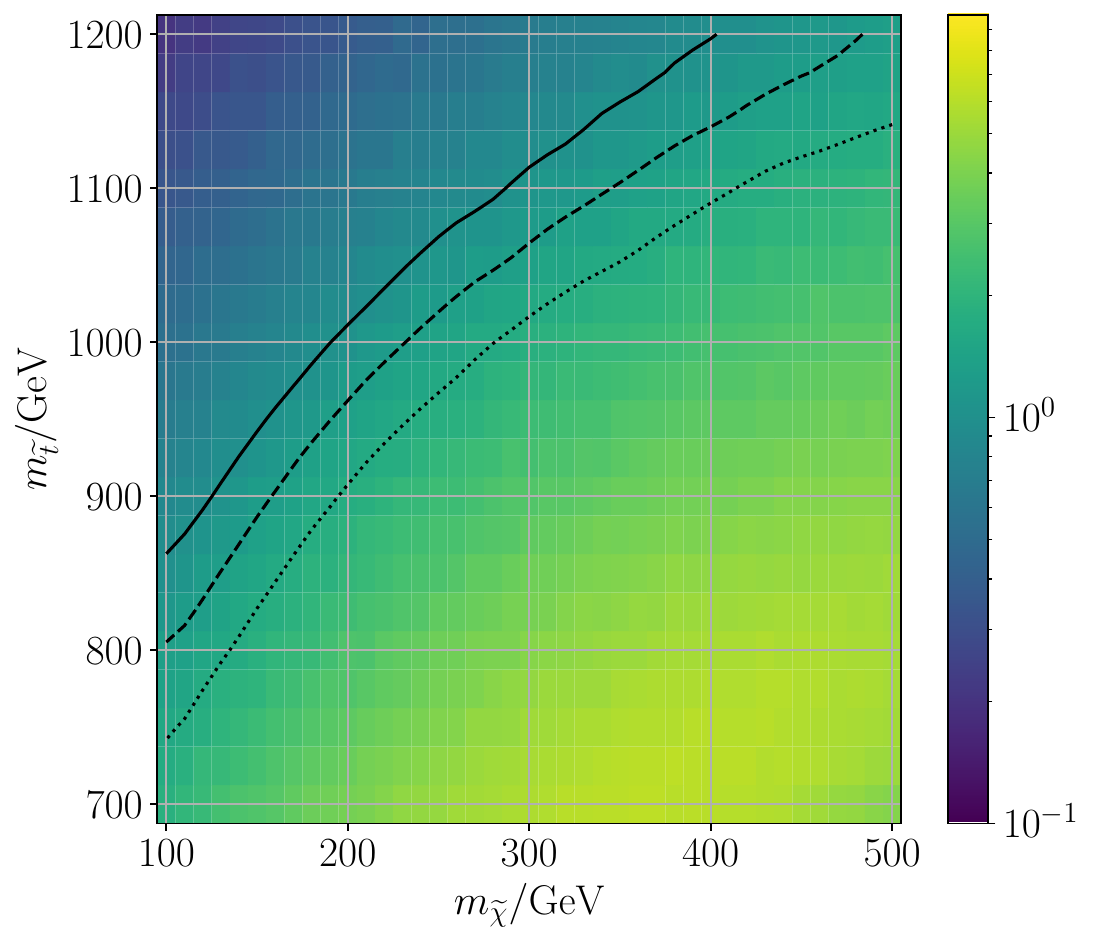}
 \includegraphics[width=0.48\linewidth]{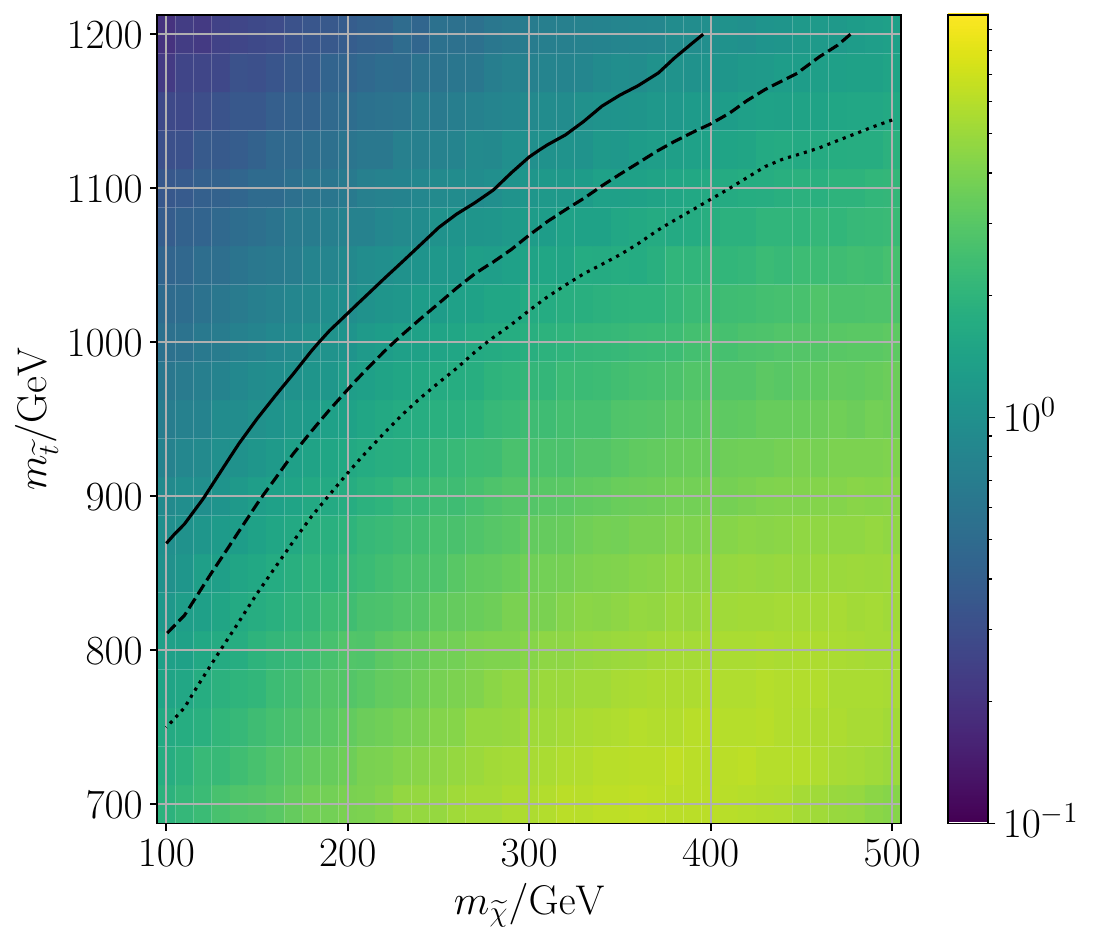}

 \includegraphics[width=0.48\linewidth]{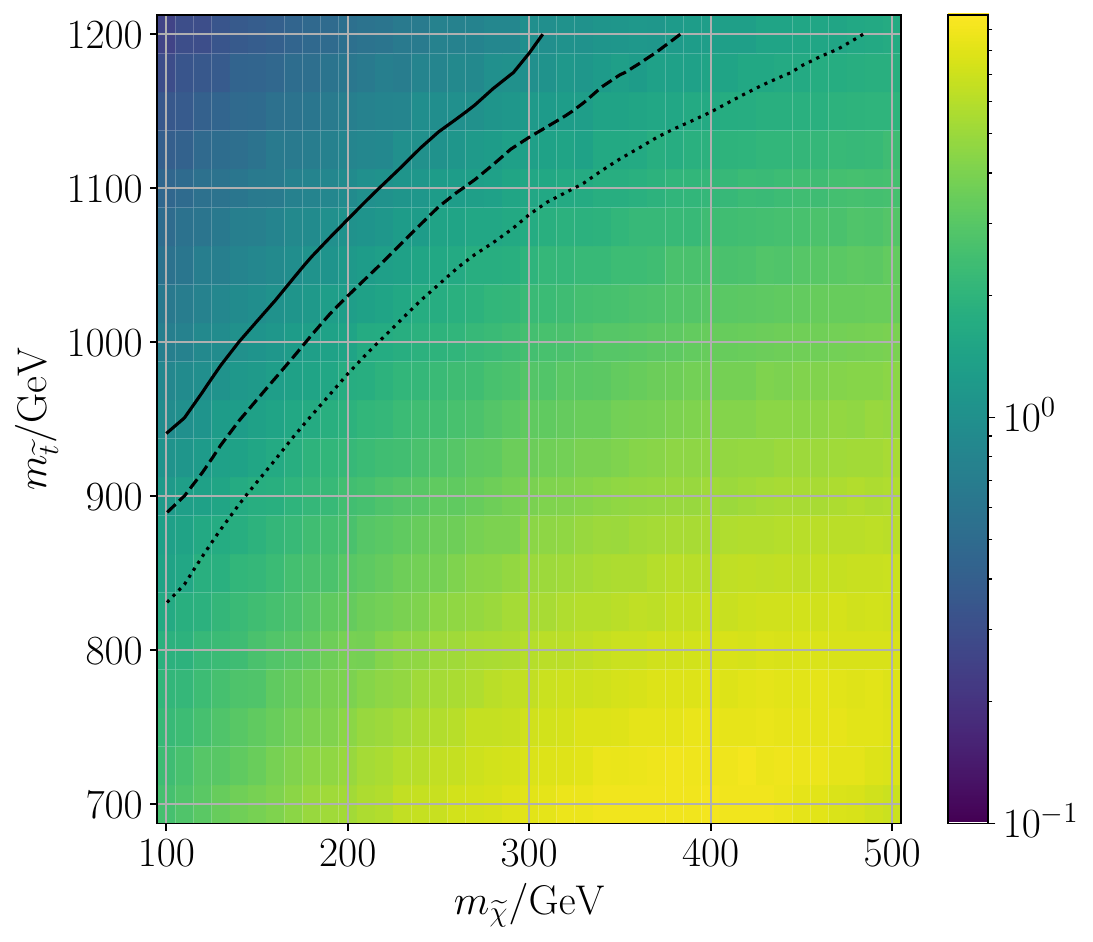}
 \includegraphics[width=0.48\linewidth]{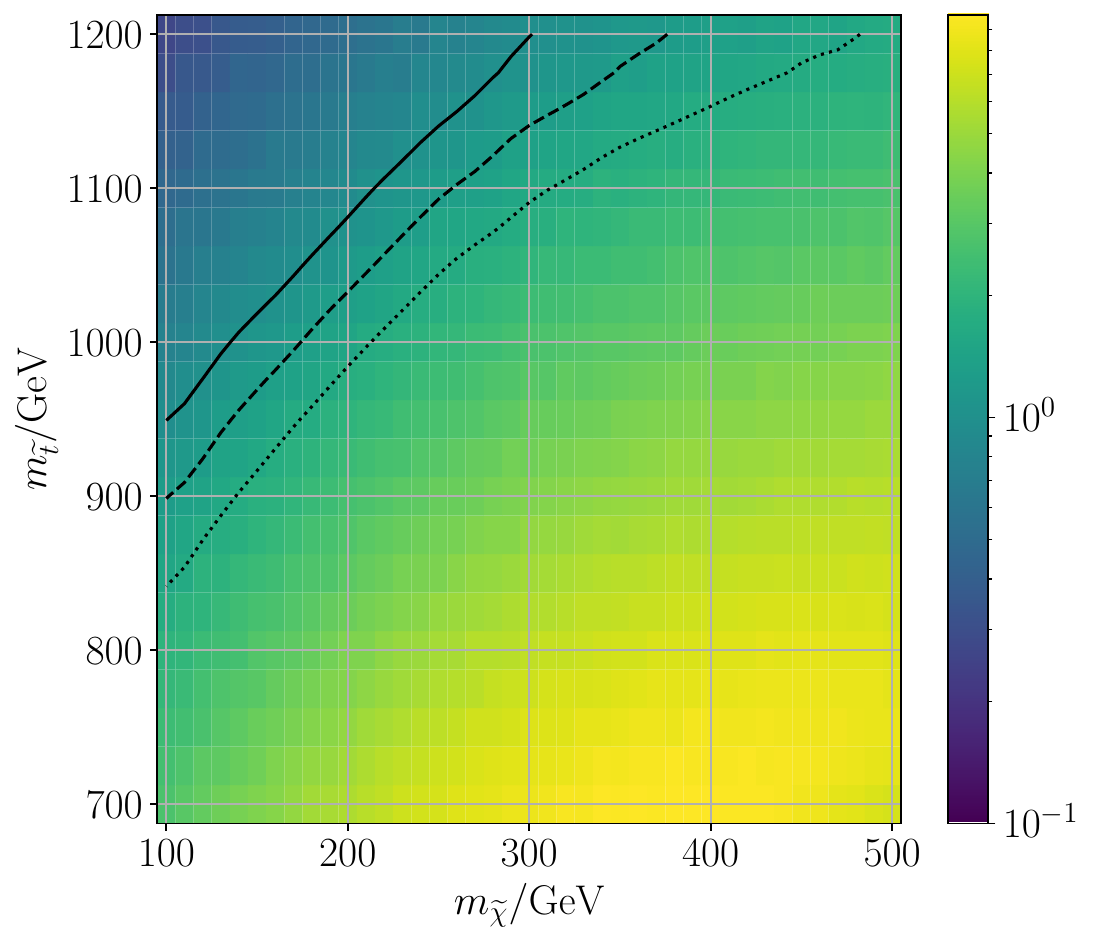}

 \caption{Exclusion significance $Z_{\rm excl}$ defined in
   eq.(\ref{eq:Zex}) for an integrated luminosity of 137
   \texorpdfstring{fb$^{-1}$}{1/fb}. In all cases the cut on the GBDT
   output has been chosen such that the signal efficiency
   $\epsilon_S=0.3$, and $\Delta_B = 0.06 B$. The top-left frame is
   for a GBDT using kinematical information only, corresponding to the
   green curve in fig.~8. The other three frames are for GBDTs that also
   use the output of LSP taggers applied to AK14 jets, based on the
   CNN (top-right), on CoAtNet (bottom-left) and on MaxViT
   (bottom-right). Solid, dashed and dotted lines denote contour lines
   corresponding to a signal significance of $1$, $1.281$ and $1.645$
   respectively. These are smoothed by a Gaussian filter with standard
   deviations $\SI{10}{\giga \eV}$ and $\SI{25}{\giga \eV}$ on the
   neutralino-mass and stop-mass axis respectively, applied to the
   logarithm of the signal significances. }
    \label{fig:exclusion137}
\end{figure*}

The results are shown in figure \ref{fig:exclusion137}. We again
define signal-like events through a cut on the GBDT output
corresponding to $\epsilon_S = 0.3$. The top-left frame is for a GBDT
that uses only kinematic information about the AK04 jets, as in the
green curve of
fig.~\ref{fig:classifiers_combined_all_mass_additional_data}.
Associating the contour along $Z_{\rm excl}=1.645$ with the $95\%$
confidence level exclusion bounds of this ``traditional'' analysis, we
find an expected exclusion reach in $m_{\tilde t_1}$ of about $740$
GeV for an LSP mass of $100$ GeV. This is rather close to the expected
reach of about $710$ GeV for the same LSP mass achieved in the CMS
search,\footnote{We note in passing that the actual CMS limit on the
  stop mass is only $670$ GeV for this LSP mass, due to a small (not
  statistically significant) excess of events.} which is based on a
neural network (NN) ``trained to recognize differences in the spatial
distribution of jets and decay kinematic distributions''
\cite{PhysRevD.104.032006}. Unfortunately they don't show results for
other LSP masses. This agreement is not accidental; we chose the
systematic background uncertainty, $\Delta_B = 0.06 B$,
accordingly. Presumably even closer agreement would have been possible
for somewhat larger $\Delta_B$. However, it would then be
significantly larger than the actual systematic error on the
background estimate found by CMS, which is below $5\%$. We note that
for $\Delta_B^2 \gg B$ the significance scales $\propto 1/B$, rather
than $\propto 1/\sqrt{B}$, if $\Delta_B$ is a fixed percentage of
$B$. A larger $\Delta_B$ therefore increases the relative improvement
in reach achieved by including information from one of our LSP
taggers; recall that this leads to a significant improvement of
$\epsilon_S/\sqrt{\epsilon_B}$, and hence to an even bigger
improvement in $\epsilon_S/\epsilon_B$.

The other three frames show results for GBDTs that also use the
outputs of an LSP tagger as input variable; we apply this tagger to
the three leading AK14 jets. We see that the simpler CNN-based tagger
(top right) increases the reach in stop mass only by less than $10$
GeV. Recall from fig.~\ref{fig:classifiers_mass} that the CNN tagger
applied on AK14 jets does not perform well for small LSP mass. For
larger LSP mass, and hence larger angular spread of the LSP decay
products, the kinematic information on the AK04 jets, many of which
are components of AK14 jets, already seems to capture much of the
physics found by the CNN. Recall that the kinematic GBDT includes
information on the angular separation of these jets via the
Fox-Wolfram moments of eq.(\ref{eq:Hl}).

In contrast, using the transformer-based LSP taggers does improve the
reach considerably. As before, MaxViT (bottom right) performs slightly
better than CoAtNet (bottom left); the reach in stop mass increases by
$100$ GeV for $m_{\tilde \chi} = 100$ GeV, and by about $60$ GeV for
$m_{\tilde\chi} = 500$ GeV. This again indicates that the kinematic
information on the AK04 jets allows some effective LSP tagging for large
LSP masses.

For stop masses in the interesting range, the
$\tilde t_1 \tilde t_1^*$ production cross section of
\cite{Borschensky_2014} can be roughly parameterized as
\begin{equation} \label{eq:sigtot}
  \sigma(pp \rightarrow \tilde t_1 \tilde t_1^*) \simeq
  0.08 \ {\rm pb} \cdot \left( \frac {m_{\tilde t_1}} {700 \ {\rm GeV}}
  \right)^{-7.8}\,.
\end{equation}
Increasing the reach from $740$ to $840$ GeV (for
$m_{\tilde\chi} = 100$ GeV) thus corresponds to reducing the bound on
the stop pair production cross section by a factor of $\sim 2.7$. Note
that the limit setting procedure is quite nonlinear, because the
background falls by nearly two orders of magnitude when
$m_{\tilde t_1}$ is increased from $700$ to $1200$ GeV while keeping
$\epsilon_S = 0.3$ fixed.

\section{Conclusion}
\label{sec:conclusion}

The large hadronic activity in $pp$ collisions makes the search for
physics beyond the Standard Model in purely hadronic processes at the
LHC especially challenging. This problem can be mitigated by the use
of sophisticated analysis methods. In particular, jet substructure has
proved a powerful discriminator between various production processes.

In this article we studied the feasibility of applying modern computer
vision techniques in detecting RPV stop decays. As a benchmark, we use
$\tilde t_1$ pair production, where each stop decays to a top and a
neutralino LSP which subsequently decays via the $UDD$ operator to
three quarks. For not too small mass splitting between the stop and
the LSP, the decay products of the latter tend to reside in a single
fat (e.g. AK14) jet. One can build images from the constituents of
such jets by using the angle $\phi$ and pseudorapidity $\eta$ as
spatial positions and deposited energy into the detector as pixel
intensity. One can then use computer vision techniques on this
representation to build classifiers (``LSP taggers'') that aid in
amplifying the signal process.

In recent years, transformer-based architectures have been shown to
trump the performance of more classical convolutional neural
network-based structures in standard classification tasks. We study
how well these novel architectures work on jet images by training LSP
taggers based on MaxViT, CoAtNet and a CNN architecture. The training
is done on single jet images. We then combine the output of the LSP
tagger applied to the three jets with the highest $p_T$ using a
gradient-boosted decision tree into a more robust classification
score. We find that the CNN-based tagger improves the statistical
significance of the signal by a factor between $5$ and $10$ for fixed
signal efficiency $\epsilon_S=0.3$, the exact factor depending on the
neutralino mass and the definition of the fat jets. In contrast, the
transformer-based models lead to an improvement factor between $8$ and
$11$, outperforming the CNN over the entire parameter space. We also
combine the predictions of all architectures for each jet size
separately and find a modest improvement, hinting that even the
transformer-based models do not use the entire information present in
the images; hence an investigation of further improvements of the
architecture might be worthwhile.

Since the kinematic preselection cuts are not optimized for
sensitivity we also use high-level features such as
Fox-Wolfram-moments, $p_T^{\rm miss}$, $H_T$, $M_J$ and $N_j$ as
inputs to a GBDT, in combination with the output of one of our LSP
taggers. This leads to a total gain of sensitivity by a factor of $20$
for $\SI{500}{\giga \eV}$ LSPs, on top of the effect due to the
acceptance cuts.

Finally, we estimate the reach in stop and LSP mass that could be
expected from the full run-2 data set. We chose the systematic
uncertainty on the background such that a GBDT that only uses
kinematic information on AK04 jets leads to a reach (for LSP mass of
$100$ GeV) similar to that found by CMS
\cite{PhysRevD.104.032006}. Using in addition the output of the
relatively simple CNN-based LSP tagger then leads to almost no further
improvement of the reach. By instead using the MaxViT-based tagger one
can improve the reach by $\SI{100}{\giga \eV}$ ($\SI{60}{\giga \eV}$)
for neutralino masses of $\SI{100}{\giga \eV}$
($\SI{500}{\giga \eV}$), under the assumption that the relative size
of the systematic uncertainty remains the same. This corresponds to
a reduction of the bound on the stop pair production cross section
by up to a factor of $2.7$.

We conclude that LSP taggers built on modern transformer-based neural
networks hold great promise in searches for supersymmetry with
neutralino LSP where $R-$parity is broken by the $UDD$ operator. This
result can presumably be generalized to models with different LSP,
e.g. a gluino decaying via the same operator, or a slepton decaying
into a lepton and three jets via the exchange of a virtual
neutralino.

In fact, it seems likely that these advanced techniques can also be
used to build improved taggers for boosted, hadronically decaying top
quarks or weak gauge or Higgs bosons. We did not attempt to construct
such taggers ourselves, since this field is already quite
mature. Convincing progress would therefore have to be based on fully
realistic detector-level simulations, for which we lack the
computational resources. Moreover, a careful treatment of systematic
uncertainties would be required, which ideally uses real
data. However, we see no reason why the improvement relative to
CNN-based taggers that we saw in our relatively simple simulations
should not carry over to fully realistic ones.

\bibliography{bibliography}
\end{document}